\documentclass[10pt,letterpaper]{article}
\usepackage[top=0.85in,left=2.75in,footskip=0.75in]{geometry}

\usepackage{amsmath,amssymb}

\usepackage[shortlabels]{enumitem}

\usepackage{changepage}

\usepackage[utf8x]{inputenc}

\usepackage{textcomp,marvosym}

\usepackage{cite}

\usepackage{nameref,hyperref}

\usepackage{xspace}

\usepackage{microtype}
\DisableLigatures[f]{encoding = *, family = * }

\usepackage[table]{xcolor}

\usepackage{array}

\newcolumntype{+}{!{\vrule width 2pt}}

\newlength\savedwidth

\raggedright
\usepackage[aboveskip=1pt,labelfont=bf,labelsep=period,justification=raggedright,singlelinecheck=off]{caption}

\makeatletter
\renewcommand{\@biblabel}[1]{\quad#1.}
\makeatother

\usepackage{lastpage,fancyhdr,graphicx}
\usepackage{epstopdf}
\fancyheadoffset[L]{2.25in}
\fancyfootoffset[L]{2.25in}
\newcommand{\WS}{Watts-Strogatz\xspace}
\newcommand{\ER}{Erd\H{o}s-R\'enyi\xspace}
\newcommand{\BA}{Barab\'asi-Albert\xspace}

\newcommand{\SIAlgorithms}{S1 Algorithms\xspace}
\newcommand{\SITable}{S1 Table\xspace}

\begin{document}
\vspace*{0.2in}

\begin{flushleft}
{\Large
\textbf\newline{Fast and exact stochastic simulations of epidemics on static and temporal networks} 
}
\newline
\\
Samuel Cure\textsuperscript{1,\dag},
Florian G. Pflug\textsuperscript{1,\dag},
Simone Pigolotti\textsuperscript{1*}
\\
\bigskip
\textbf{1} Biological Complexity Unit, Okinawa Institute of Science and Technology, Onna, Okinawa 904-0495, Japan.
\\
\bigskip

%
%
%
\dag These authors contributed equally to this work.





* simone.pigolotti@oist.jp

\end{flushleft}
\section*{Abstract}
Epidemic models on complex networks are widely used to assess how the social structure of a population affects epidemic spreading. However, their numerical simulation can be computationally heavy, especially for large networks. In this paper, we introduce NEXT-Net: a flexible implementation of the next reaction method for simulating epidemic spreading on both static and temporal weighted networks. We find that NEXT-Net is substantially faster than alternative algorithms, while being exact. It permits, in particular, to efficiently simulate epidemics on networks with millions of nodes on a standard computer. It also permits simulating a broad range of epidemic models on temporal networks, including scenarios in which the network structure changes in response to the epidemic. NEXT-Net is implemented in C++ and accessible from Python and R, thus combining speed with user friendliness. These features make our algorithm an ideal tool for a broad range of applications.

\section*{Author summary}

Human social structures tend to be quite heterogeneous, with some individuals having many more social interactions than others. These social structures profoundly affect the spreading of epidemics and can be conveniently conceptualized as networks, in which nodes represent individuals and links represent contacts. However, computer simulations of epidemic models on networks can be slow, and efficient numerical methods are understudied. This prevents computer simulations of epidemics on realistically large networks. In this paper, we present NEXT-Net: an algorithm to efficiently simulate epidemic spreading on networks. Our algorithm can simulate a broad class of models, including networks whose structure evolves in time.  Its versatility, ease of use, and performance make it broadly useful for  epidemiological studies.






\section*{Introduction}

Mathematical models are invaluable tools to rationalize the spreading of epidemics. The simplest models assume that epidemics spread in well-mixed populations \cite{grassly2008mathematical}. However, this simplifying assumption neglects fundamental aspects such as the heterogeneity of contacts in a population and the presence of social structures. A common and powerful way to include these factors is to model epidemic spreading as a process taking place on a network \cite{pastor2015epidemic,kiss2017mathematics,moore2020predicting,cure2024rate}, where nodes represent individuals and links represent contacts. In this class of models, infected individuals can infect their contacts according to certain stochastic rules. Computer simulations of these models are, unfortunately,  computationally demanding on large networks \cite{pastor2015epidemic}. Efficient numerical methods are thus crucial.

In an epidemic, the infectiousness of individuals, i.e., their propensity to spread the disease, depends on the time since they were infected in a disease-specific manner \cite{grassly2008mathematical}. This time dependence strongly affects epidemic spreading and therefore has to be taken into account in models. 
In the literature, models with time-dependent rates are often referred to as ``non-Markovian'' \cite{grassly2008mathematical,pastor2015epidemic}. Non-Markovian models can not be simulated using the standard
Gillespie algorithm \cite{gillespie1977exact}.

Several algorithms for simulating non-Markovian epidemic models on networks have been proposed 
\cite{boguna2014simulating,masuda2018gillespie,pelissier2022practical}. Each possesses its own advantages and disadvantages. The non-Markovian Gillespie Algorithm (nMGA) \cite{boguna2014simulating} generalizes the Gillespie algorithm to arbitrary infection time distributions. However, the time it takes for nMGA to process a single infection scales linearly with the number of infected nodes, making simulations infeasible for large networks. In addition, the nMGA is exact only in the limit of a large number of infected nodes. The Laplace Gillespie algorithm \cite{masuda2018gillespie} is exact and has a lower computational complexity than nMGA. However, it can only be used for monotonically decreasing infection time distributions, which excludes most realistic cases. The Rejection-based Gillespie for non-Markovian Reactions (REGIR) algorithm \cite{pelissier2022practical} efficiently simulates epidemics for arbitrary distributions, but, like nMGA, it is an approximate algorithm. Finally, the next reaction method is a flexible and exact algorithm, originating from chemical physics \cite{gibson2000efficient,anderson2007modified}, that has been applied to simulate epidemic spreading \cite{kiss2017mathematics,feng2019equivalence}. However, it has been doubted whether it can be used effectively for large networks \cite{pastor2015epidemic}. 

The algorithms we mentioned can be used to simulate epidemic spreading on static networks. However, the social structures that affect epidemic spreading may evolve over time, thus requiring a description in terms of temporal networks \cite{holme2012temporal}. The network evolution can be independent of the disease, or can arise as a response to the disease itself. For example, infected individuals may behave differently than susceptible ones \cite{wang2021impact} and the whole population may change its behavior as the disease spreads \cite{paarporn2017networked,hu2018epidemic}. Efficient numerical algorithms to study epidemic spreading on temporal networks have received little attention so far.

In this paper, we present NEXT-Net (Next-reaction-based Epidemics eXtended to Temporal NETworks), a simulation algorithm for epidemics on both static and temporal weighted networks. Our algorithm is based on a combination of the next reaction scheme and rejection sampling to handle temporal networks efficiently. By a systematic comparison in the static network case, we find that NEXT-Net is much faster than alternatives (nMGA and REGIR) in all the examples we considered. In particular, NEXT-Net scales approximately linearly with the network size, thus allowing efficient simulations of epidemic models on networks with millions of nodes on a standard computer. Our algorithm is implemented in C++ for performance, and accessible from both Python and R for ease of use. 

\begin{figure}[!ht]
    \centering
    \includegraphics[width=\linewidth]{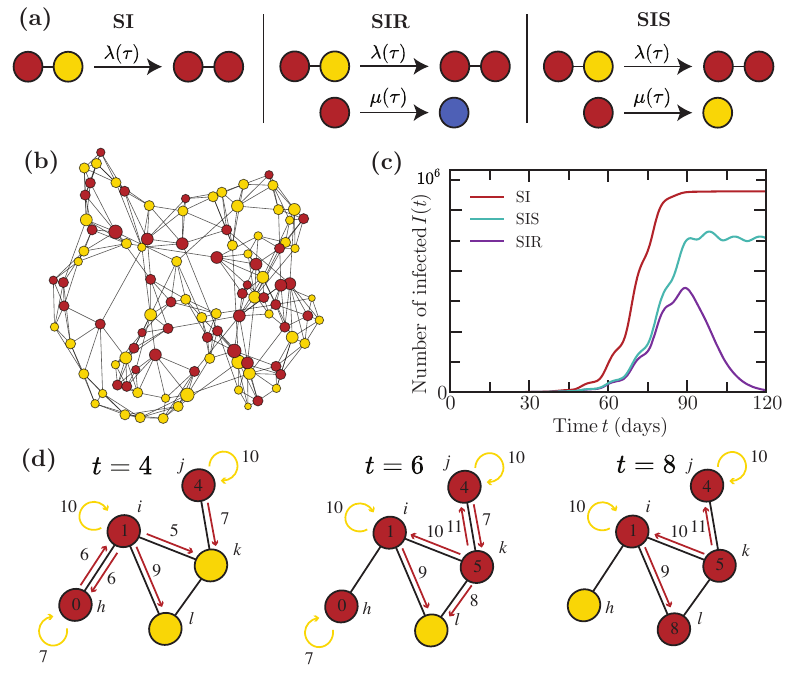}
    \caption{{\bf Epidemics on  networks}. $(a)$ The SI, SIR, and SIS models. Susceptible nodes are represented in yellow, infected nodes in red, and recovered (immune) nodes in blue. $(b)$ A state of an epidemic on a Watts-Strogatz network \cite{watts1998collective}. $(c)$ Average epidemic trajectories on Watts-Strogatz networks of size $n=10^5$ for infection times that are Gamma distributed with mean 5 and variance 3 and recovery times that are Gamma distributed with mean 12 and variance 5. Oscillations in the trajectories appear due to the shape of these functions. $(d)$ Illustration of the next reaction method for the SIR model.  Numbers on the nodes represent the time $t$ at which they contracted the infection. The times $t$ on the red arrows are the times at which nodes transmit the infection via a given link. We assign transmission times even if a link connects two infected nodes, in which case transmission has no effect. Numbers on the yellow arrows represent the recovery times. }
    \label{fig:1_next_reaction}
\end{figure}

\section*{Models and Algorithms}

\subsection*{Epidemics on static networks}

We introduce non-Markovian epidemic models on static weighted networks. Network nodes $i=1\dots N$ represent individuals, who can be in a susceptible (S), infected (I), and possibly recovered (R) state. A link from node $i$ to node $j$ represents a contact along which infected individual $i$ can spread the disease to $j$. We assume in general directed networks, although several of our examples will be non-directed. Links are assigned  weights $w_{ij}\in[0,\infty)$. We consider the three classic models (Fig.~\ref{fig:1_next_reaction}a):

\begin{description}
\item[Susceptible-Infected (SI).] Infected individuals transmit the disease to their susceptible contacts at a rate $w_{ij}\lambda(\tau)$, where $\tau$ is the time since their infection. The unweighted case is recovered by setting $w_{ij}=1$ for all links $(i,j)$. We call $\psi(\tau\,|\,w_{ij})$ the probability density of transmitting the disease at time $\tau$ along a given link. This density is related with the spreading rate by
\begin{equation}\label{eq:psi_static}
    \psi(\tau\,|\,w_{ij})=w_{ij}\lambda(\tau)\exp\left(-w_{ij}\int_0^\tau \lambda(\tau')d\tau'\right) .
\end{equation}
Any distribution can be written in the form of Eq.~\eqref{eq:p_psi_static} by a suitable choice of $\lambda(\tau)$, see \SIAlgorithms{} for details. The probability that an infected node eventually transmits the disease along a link having unit weight is given by
\begin{equation}\label{eq:p_psi_static}
    p_\psi=\int_0^\infty \psi(\tau)d\tau = 1 - \exp\left(-\int_0^\infty \lambda(\tau)d\tau\right),
\end{equation}
For arbitrary weight, the probability of eventual transmission is $1 - (1-p_\psi)^{w_{ij}}$. For quickly decaying $\lambda(\tau)$ such that $p_\psi < 1$, the distribution $\psi(\tau\,|\,w_{ij})$ is thus not normalized.

\item[Susceptible-Infected-Recovered (SIR).] In this extension of the SI model, infected individual can recover, and recovered individuals can neither transmit the disease nor be reinfected. Recovery occurs at a time-dependent rate $\mu(\tau)$, leading to a distribution of recovery times
\begin{equation}\label{eq:rho_static}
\rho(\tau)=\mu(\tau)\exp\left(-\int_0^\tau\mu(\tau')d\tau'\right)
\end{equation}
and a probability of eventual recovery of $p_\rho=\int_0^\infty \rho(\tau)d\tau = \exp\left(-\int_0^\infty \mu(\tau) d\tau\right)$. In well-mixed populations, the SIR model is equivalent to an SI model with modified infectiousness $\widetilde \lambda(\tau) = \lambda(\tau)\exp\big(-\int_0^\infty \mu(\tau') d\tau'\big)$ where the exponential factor represents that probability that an individual has not recovered. On networks, however, this equivalence is no longer exact: recovery times are assigned to nodes rather than links, and transmission times thus become correlated once recovery is taken into account. We therefore simulate recovery as a separate event and discard transmissions which would take place after a node has recovered.

\item[Susceptible-Infected-Susceptible (SIS).] In this variant of the SIR model, recovery makes individuals  susceptible again. For simplicity of implementation, we assume that each infected individual can infect each of their neighbors at most once before recovering. This does not preclude individuals spreading the disease to their neighbors multiple times if they contract the disease repeatedly.

\end{description}

These three models can be defined on an arbitrary static network (Fig.~\ref{fig:1_next_reaction}b) and produce markedly different epidemic trajectories (Fig.~\ref{fig:1_next_reaction}c).

\subsection*{Epidemic models on temporal networks}

We now extend the SI, SIR, and SIS models to temporal networks, i.e., networks in which links are created and removed at certain moments in time. We represent a temporal network as a field of functions $\varepsilon_{ij}(t)$, whose value is one if a link between node $i$ and $j$ exists at time $t$ and zero otherwise.  The network evolution can be deterministic or stochastic. In particular, $\varepsilon_{ij}(t)$ may depend on the epidemic state of the nodes up to time $t$.

An infected individual $i$ can infect $j$ at time $t$ only if a link between $i$ and $j$ exists at time $t$. This means that the effective transmission rate between $i$ and $j$ is now $w_{ij}\lambda(\tau)\varepsilon_{ij}(t)$, where $\tau$ is the time since infection of $j$. It follows that the effective distribution of infection times from node $i$ to node $j$ is
\begin{equation}\label{eq:psi_temp}
\psi_{i,j}(\tau\,|\,w_{ij}\,;\,T_i)=w_{ij}\lambda(\tau)\varepsilon_{ij}(T_i+\tau) \exp\left(-w_{ij}\int_0^\tau \lambda(\tau')\varepsilon_{ij}(T_i+\tau')d\tau'\right)\, ,
\end{equation}
where $T_i$ is the absolute time at which individual $i$ was infected. Equation~\eqref{eq:psi_temp} is the equivalent of Eq.~\eqref{eq:psi_static} for temporal networks. We observe that, in contrast with the static network case, $\psi_{ij}$ now depends both on the time since infection of node $i$ and on the absolute time $t$ (through $T_i$).  For SIR and SIS models, the recovery distribution $\rho(\tau)$ is defined as for static networks, see  Eq.~\eqref{eq:rho_static}.

In some real temporal networks, contacts are brief enough to be treated as instantaneous, with a non-negligible chance of infection in each contact. Mathematically, such contacts can be represented as Dirac $\delta$ peaks in $\varepsilon_{ij}(\tau)$ with weight $w_{ij}$. In this case $\psi_{i,j}(\tau\,|\,w_{ij}\,;\,T_i)$ becomes a probability distribution over a set of discrete events. The individual transmission probability during each such instantaneous contact, assuming that no earlier transmission took place, is given by $1 - \exp(-w_{ij}\lambda(\tau))$. 

\subsection*{Simulation algorithms}

We now examine algorithms for simulating the spread of epidemics on static and temporal networks.  We do not consider the Laplace-Gillespie algorithm, since it can only be used if the  infection time distribution is monotonically decreasing, which is not the case for most diseases. 

\subsubsection*{Next Reaction Method (NEXT-Net)}

We here describe the implementation of the next reaction method in NEXT-Net, see Fig.~\ref{fig:1_next_reaction}d. Every time an individual is infected, we draw the times until infection of each neighbor from the distribution $\psi(\tau\,|\,w_{ij})$. For SIR and SIS models, we also draw the random time until recovery from the distribution $\rho(\tau)$. The absolute times of these events, together with their type (infection or recovery) and the participating nodes, are inserted into a global priority queue. At each step of the algorithm, we retrieve the earliest event from this queue and execute it. In the case of infections, this operation on average adds $R_0$ further future events into the queue, where $R_0$ is the basic reproduction number (i.e. average number of subsequent infections caused by a single infection). The resulting time complexity of a single step when $I$ nodes are infected (and the size of the priority queue is thus at most $I R_0$) is dominated by the complexity of maintaining the priority queue, that is at most $\mathcal{O}\big(\log(I R_0)\big)$. The algorithm is described in detail in \SIAlgorithms{}.

\subsubsection*{non-Markovian Gillespie (nMGA)}

The non-Markovian Gillespie algorithm (nMGA) \cite{boguna2014simulating} extends the Gillespie algorithm to time-varying infectiousness functions $\lambda(\tau)$ by neglecting variations in $\lambda(\tau)$ between subsequent global events. The cumulative distribution $\Phi(\tau)$ of the time until the next event is thereby approximated by the exponential distribution
\begin{equation}\label{eq:ngma_phi}
    \Phi(\tau)\approx \exp\left(-\tau \sum_{i,j}{w_{ij}\lambda(\tau_i)}\right).
\end{equation}
where $i$, $j$ ranges over all links such that node $i$ is infected and $\tau_i$ denotes the time since infection of node $i$. The algorithm tends to be exact when the number of infected individuals is very large, since the time between events tends to zero in this limit. However, since $\lambda(\tau_j)$ must be evaluated for every infected individual, a single time step has time complexity $\mathcal{O}(I R_0)$.

\subsubsection*{Rejection-based Gillespie for non-Markovian Reaction (REGIR)}

The REGIR algorithm \cite{pelissier2022practical} is an optimized version of the nGMA algorithm for unweighted networks in which $\lambda(\tau)$ in Eq.~\eqref{eq:ngma_phi} is replaced by an upper bound $\lambda_{\rm max} \geq \sup_\tau {\lambda(\tau)}$. The resulting under-estimation of the time until the next event is then corrected by accepting events at time $\tau$ with probability $\lambda(\tau) / \lambda_{\rm max}$.  This modification eliminates the need to evaluate $\lambda(\tau_i)$ for each infected node $i$ and thus reduces the time complexity of a single time step down to $\mathcal{O}(1)$. However, the time required for a single step is inversely proportional to the acceptance rate. The advantage of REGIR over nGMA thus depends on the choice of $\psi(\tau)$ and may be small if $\lambda(\tau)$ is characterized by narrow peaks. Despite the formally lower time complexity of $\mathcal{O}(1)$ for REGIR vs. $\mathcal{O}\big(\log(I R_0)\big)$ for NEXT-Net, we shall see that NEXT-Net is considerably faster in practice, see Results.

\subsubsection*{NEXT-Net for temporal networks}

NEXT-Net extends the next reaction method to simulate epidemics on temporal networks. It is designed to only require information on the network up to the present time and is therefore apt to simulate temporal networks whose structure evolves in response to the epidemics. This feature prevents us from simply mapping the temporal case into the static case by means of Eq.~\eqref{eq:psi_temp}.  NEXT-Net evolves the network in lock-step with the epidemics.  At every time step, we query two times: (i) the tentative next time a link is added or removed, and (ii) the tentative next time when a node is infected or recovers, and execute the earlier event.

\begin{figure}[!ht]
    \centering
    \includegraphics[width=\linewidth]{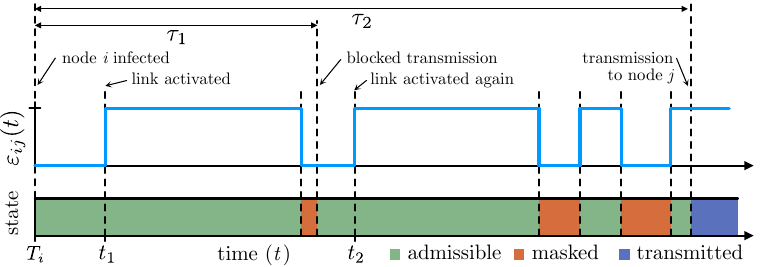}
    \caption{{\bf Epidemics on temporal  networks}. Illustration of the temporal simulation algorithm. A link from node $i$ to node $j$ appears and disappears after node $i$ has been infected at time $T_i$. Upon appearing at time $t_1$, the link is ``activated'', i.e., a time $\tau_1 \geq t_1 - T_i$ between infection and transmission is drawn from $\psi(\tau\,|\,w_{ij})$, where the condition $\tau_1 \geq t_1 - T_i$ ensures that the time lies in the future, i.e. after $t_1$. However, before the simulation reaches time $T_i + \tau_1$, the link disappears, $\varepsilon_{ij}(t)=0$, causing the algorithm to mask the link. Since the link is still masked when transmission is attempted at time $T_i + \tau_1$, the attempt is blocked. The link is then unmasked so that when it reappears at time $t_2$, it is re-activated, i.e. a time until transmission $\tau_2$ (again conditioned to lie in the future, i.e. after $t_2$) is drawn. Further disappearances and reappearances of the link before time $t = T_i + \tau_2$ then do not cause further activations but merely change the state of the link. Once the simulation reaches time $t = T_i + \tau_2$, the disease is then transmitted to node $j$ since the link happens to be unmasked at that time. See \SIAlgorithms{} for a detailed description of the algorithm.}
    \label{fig:temporal_networks}
\end{figure}

To generate transmission times distribution according to Eq.~\eqref{eq:psi_temp} without knowledge of the future evolution of $\varepsilon_{ij}$, we employ a rejection sampling scheme (Fig.~\ref{fig:temporal_networks}). When a node is infected, it is initially handled as in the static case. For each neighbor present, at the time of infection the link is ``activated'', i.e. an infection time with distribution $\psi(\tau)$ is generated. This step tentatively assumes the link will remain present until transmission occurs. If new links are later added to already infected nodes, these are immediately activated as well, but the transmission time is generated assuming no transmission prior to the link's appearance, see \SIAlgorithms for details. If a link connected to an infected node is removed, we mark this link as ``masked'', which blocks subsequent transmission events. If a masked link attempts to transmit and it is later re-added, it is treated as a new link. If a masked link is re-added before its transmission time, it is simply unmasked. To distinguish these two cases, masked links are unmasked upon an attempted transmission. This masking/unmasking scheme effectively employs ``thinning'' \cite{lewisSimulationNonhomogeneousPoisson1979} to sample the first firing time of a Poisson process with intensity $w_{ij}\lambda(\tau)\varepsilon_{ij}(T_i + \tau)$ from the firing times generated for a process with intensity $w_{ij}\lambda(\tau)$. See \SIAlgorithms for a detailed description and pseudo-code of the algorithm.

NEXT-Net also allows for instantaneous contacts, that simply transmit the disease with probability $1 - \exp(-w_{ij}\lambda(\tau))$. The algorithm described in this Section can, in principle, be used together with any simulation algorithm for static networks. In our implementation it is, however, currently restricted to the next reaction method.

The contribution of the temporal NEXT-Net algorithm to the time complexity of generating a single infection is $\mathcal{O}\big(K\log(I R_0)\big)$, assuming $K$ transmission attempts until an unmask link is encountered on average, where $R_0$ is now the average number of simultaneous neighbors (see \SIAlgorithms). In practice, however, we observe that the computational bottleneck is usually caused by the network evolution. The complexity of this step depends on the specific temporal network model.

\subsection*{Implementation}

We implemented three algorithms (NEXT-Net, nMGA, REGIR) in C++ to ensure maximal performance and made them accessible from both Python and R for convenience. 

Our implementations do not presume any specific networks or infection/recovery time distributions and support SI, SIR, and SIS models. For convenience, we provide a range of classic network models (such as \WS \cite{watts1998collective}, \ER, and \BA \cite{newman2018networks}) among others and allow arbitrary weighted networks defined in terms of an adjacency list or edge list to be used. Infection/recovery time distributions can either be specified by specifying $\lambda(\tau)$ through vectors $(\tau_i)$, $(\lambda_i)$, or by classic infection time distributions like exponential, Gamma, Lognormal, and Weibull, see \SIAlgorithms for a detailed list of options. Users can also easily implement their own networks and time distributions through flexible interfaces. When used from Python, our code also allows seamless access to all network models available in NetworkX \cite{hagberg2020networkx}.

\section*{Results}

\subsection*{Epidemics on static networks}
\begin{figure}[htbp]
    \centering
    \includegraphics[width=\textwidth]{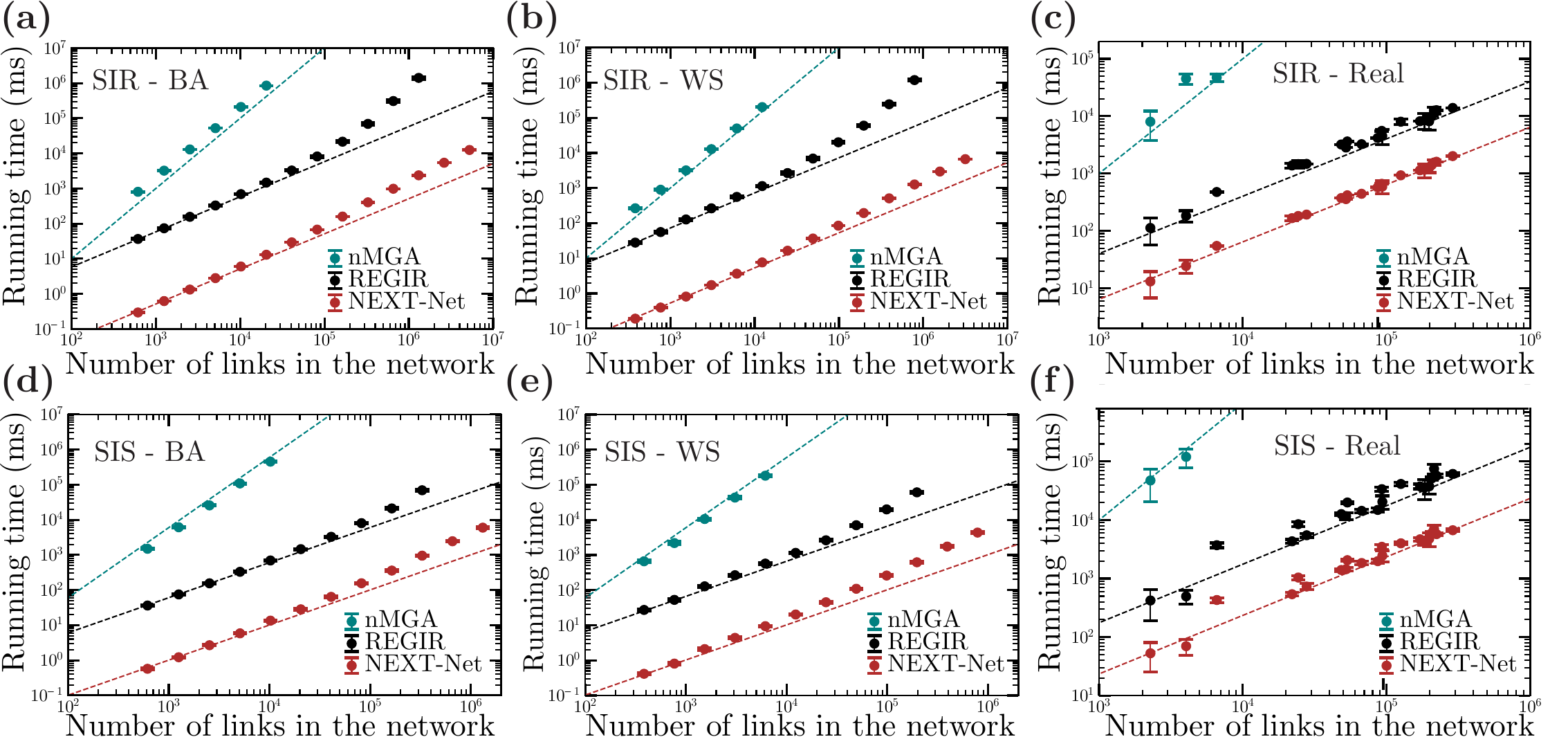}
    \caption{{\bf Numerical test for simulations of epidemics on various networks.} We simulate SIR: \textbf{(a)}, \textbf{(b)}, \textbf{(c)} and SIS: \textbf{(d)}, \textbf{(e)}, \textbf{(f)} epidemic processes using various algorithms on various networks: on \WS networks \textbf{(a)},\textbf{(d)}; \BA networks \textbf{(b)},\textbf{(e)} and on real-world networks \textbf{(c)},\textbf{(f)}.  A list of the real world networks, their parameters, and their mean simulation times are reported in \SITable. The infection times are Gamma distributed with mean $5$ and variance $3$ while the recovery time are Gamma distributed with mean $10$ and variance $12$. For each network we repeat the simulations 100 times. Dots represent  average times and bars represent standard deviations. Simulations were executed on a workstation with an Intel\textregistered{} Xeon\textregistered{} 6128 CPU @ 3.40\,GHz under Ubuntu 24.04.}
    \label{fig:numerical_comparison}
\end{figure}

We simulated SIR and SIS models on different static  networks, and found that our implementation of the next reaction method in NEXT-Net consistently outperforms both nMGA \cite{boguna2014simulating} and REGIR \cite{pelissier2022practical} (Fig.~\ref{fig:numerical_comparison}). In these comparisons, we used the paradigmatic \BA and \WS models, and also real-world networks from different databases \cite{clauset2016colorado,snapnets,konect}. To ensure a fair comparison, we use a Gamma distribution for transmission and recovery times. This distribution allows for an efficient bound $\lambda_\text{max}$, which favors the REGIR algorithm. However, our benchmark indicates that REGIR still seems to spend more time on retries than NEXT-Net does to maintain its priority queue. Simulations are always initialized with a single infected node, chosen at random. For SIR epidemics, we measure the average time to simulate an SIR epidemic until no infected nodes remain. For SIS epidemics, we stop the simulation at a time $T_{\rm max}$, chosen as the average time for the SIR epidemic to end on the same network.

The runtime for both NEXT-Net and REGIR appears to scale slightly super-linearly. For NEXT-Net, a possible explanation may be the logarithmic dependency of the runtime on the size of the priority queue. However, since REGIR shows super-linear scaling as well, it is also possible that the working set starts to exceed the cache size at around $10^5$ nodes. Despite presenting similar scaling, the NEXT-Net outperforms the REGIR algorithm by roughly a factor of 10 in speed. For example, for a SIR epidemic on a \WS network of size $2.6\times 10^5$, the average time to complete an epidemic for REGIR is 24 minutes 50 seconds, while NEXT-Net only takes 4.7 seconds. We expect this performance gap to be even larger for other infection time distributions. Finally,  the total runtime of nMGA appears to scale quadratically with the total number of links in the network, as we would expect from its computational complexity. This makes nMGA substantially slower than both REGIR and NEXT-Net, preventing simulations on large networks. 

\begin{figure}[ht]
    \hbox to \textwidth{
    \vtop{\hbox to 0in{\ \textbf{(a)} $\quad$  $\quad$ $\quad$  $\quad$   \WS }\hbox{\includegraphics[scale=0.5]{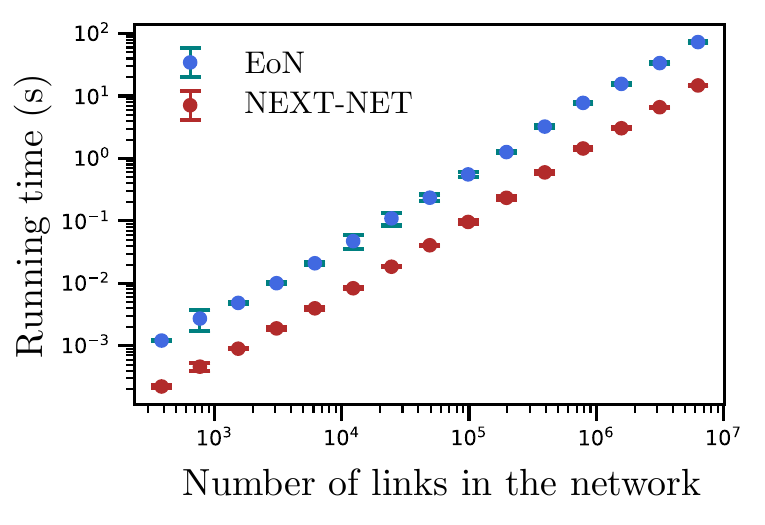}}}%
    \hfill
    \vtop{\hbox to 0in{\ \textbf{(b)} $\quad$  $\quad$   $\quad$  $\quad$ \BA }\hbox{\includegraphics[scale=0.5]{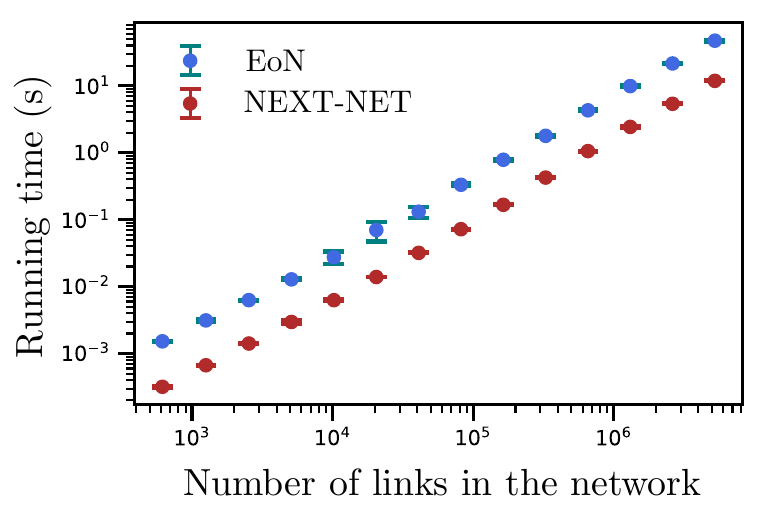}}}
    }
    \caption{{\bf Comparison of performance between our next reaction implementation and the Python library from Ref.~\cite{kiss2017mathematics}.} We simulate SIR epidemic processes on \WS ~networks \textbf{(a)} and \BA ~networks \textbf{(b)} using the Python wrapper of our C++ implementation and compare its performance with the Python library from Ref.~\cite{kiss2017mathematics} (EoN). The infection times are Gamma distributed with mean $5$ and variance $3$ while the recovery time are lognormal with mean $10$ and variance $12$. For each network we repeat the simulations 100 times, the dots represent the average time and the bars represent the standard deviation. }
    \label{fig:numerical_comparison_eon}
\end{figure}

We also compare our implementation on static networks with a Python library implementing a next reaction method for epidemics on networks \cite{kiss2017mathematics}, see Fig.~\ref{fig:numerical_comparison_eon}. As expected, we obtain a similar scaling in computational complexity since both implementations use a priority queue. However, NEXT-Net is about one order of magnitude faster, likely because it is implemented in C++.

\subsection*{Epidemics on temporal networks}

We now demonstrate the use of NEXT-Net for epidemic simulations on temporal networks. In this case, we are not aware of other established algorithms to compare with. We consider epidemics spreading on three different types of temporal networks: (1) An activity-driven network in which nodes randomly activate and deactivate, affecting their connectivity. (2) A network defined by spatial proximity of diffusing particles. (3) Empirically observed networks consisting of instantaneous contacts between nodes.

\subsubsection*{Activity-driven networks}

In an activity-driven network model, nodes stochastically alternate between an active and inactive state. Nodes lose all of their links when they are inactivated and form new connections upon activation. We here focus on a specific model inspired from Ref.~\cite{cai2024epidemic}. The network model is defined as follows. Inactive nodes becomes active at a constant rate $a$, and active nodes deactivate at constant rate $b$. Upon activation, a node connects to $m$ other nodes, selected uniformly at random and not necessarily active. Upon deactivation of a node, all its links are severed. Before starting an epidemic on such networks, we simulate the network dynamics until a steady-state is reached. 

\begin{figure}[htbp]
    \centering
    \includegraphics[width=\textwidth]{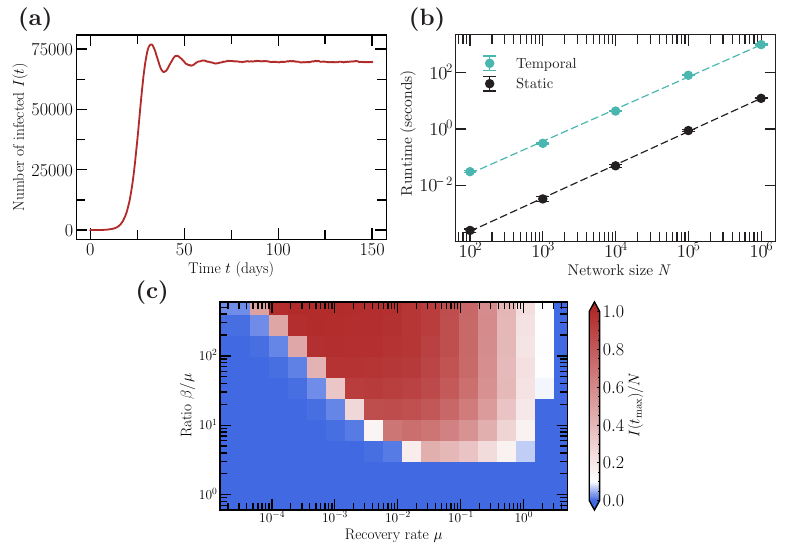}
    \caption{{\bf Epidemics on an activity driven network.} Panel (a): A SIS epidemic on an activity driven network of size $N=10^5$ with activation rate $a=0.1$, deactivation rate $b=0.1$, and $m=3$. The infection times are Gamma distributed with mean $3$ and variance $1$, the recovery times are lognormally distributed with mean $10$ and variance $1$. Panel (b): Runtime for a SIR epidemic on an activity driven network as a function of the network size. We average over 100 simulations. Panel (c): Phase diagram of the SIS model for a constant infection and recovery rates $\beta,\mu$ on an activity driven network of size $10^5$. Each simulation is initialized with the degree distribution at equilibrium and ends at a time $t_{\max}$ when the epidemic is at steady state. }
    \label{fig:example-activity-driven}
\end{figure}

We run SIR and SIS models on such an activity network, see Fig.~\ref{fig:example-activity-driven}a. The epidemic is seeded with a single infection after the activity dynamics has reached a steady state, as indicated by a constant average degree. The computational time scales approximately linearly with the network size as expected, see Fig.~\ref{fig:example-activity-driven}b. For moderately large network size ($N=10^5$), approximately 62\% of computational time is devoted to the activity dynamics, 31\% for simulating the epidemic,  and the remaining 7\% to notify the epidemic process of the appearance of new active links. This means that the main computational cost is due to updating the temporal network, while the epidemic algorithm is rather efficient. To confirm this, we measure the average time it takes to run an epidemic on an equivalent static network. When the activity driven network is in equilibrium, there are on average $N \langle k\rangle /2$ links at any given time where $\langle k \rangle =m \left((a+b)^2+b^2\right)/(a+b)^3$ \cite{cai2024epidemic}. In our simulations this gives $\langle k \rangle=0.27$. A static network with the same average degree would on average not support an epidemic outbreak due to the lack of a giant component. We thus consider a \ER ~network with $\langle k \rangle=10$ to ensure an exponential outbreak. As expected, the computational time on these static networks is much lower than the temporal ones, see Fig.~\ref{fig:example_empirical_temporal_network}b.


Epidemic spreading on a temporal network drastically differs from the case of static networks when the network dynamics and the epidemic operate on a comparable time scale. As an example, we simulated the SIS model on this activity network for different values of the infectiousness and recovery rates, where $\psi(\tau)$ and $\rho(\tau)$ are exponential distributions with rate $\beta$ and $\mu$. On static networks, the epidemic threshold is a function of $\beta/\mu$ only. In contrast, here the epidemic threshold does not only depend on their ratio, but also on the timescale of recovery, see Fig.~\ref{fig:example-activity-driven}c, in agreement with the results of Ref.~\cite{cai2024epidemic}. 

\subsubsection*{Brownian proximity networks}

\begin{figure}[htbp]
    \hbox to \textwidth{
    \vtop{\hbox to 0in{\ \textbf{(a)}}
        \hbox{\includegraphics[scale=0.33]{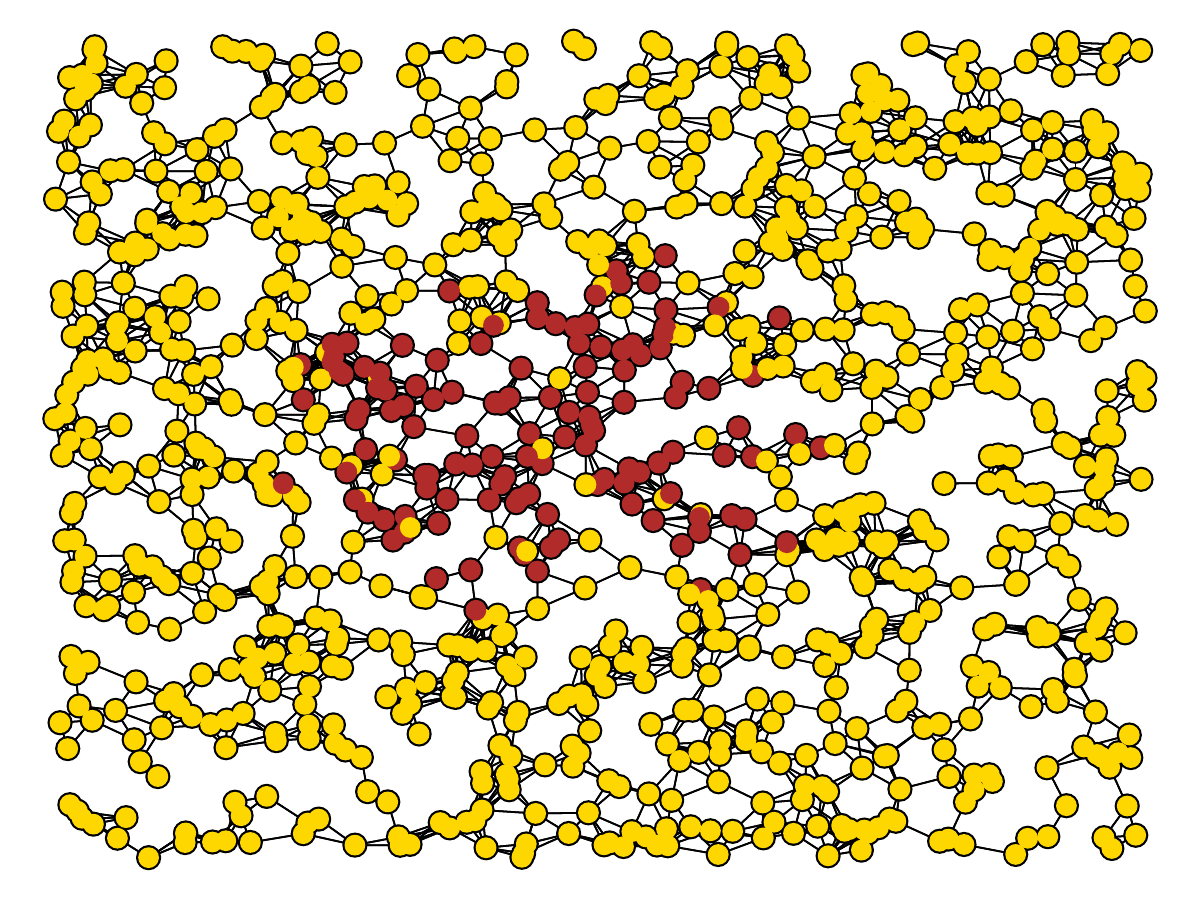}}}%
    \hfill
    \vtop{\hbox to 0in{\ \textbf{(b)}}
        \hbox{\includegraphics[width=0.50\textwidth]{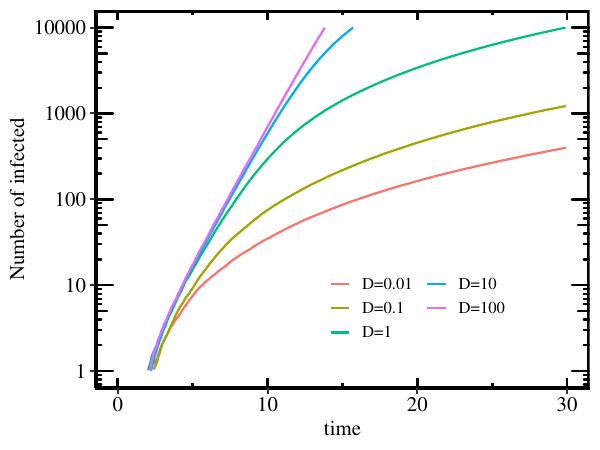}}}%
    }
    \hbox to \textwidth{
    \vtop{\hbox to 0in{\ \textbf{(c)}}
        \hbox{\includegraphics[width=0.30\textwidth]{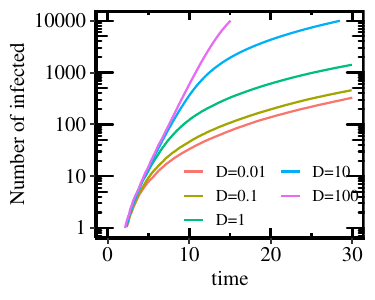}}}%
    \hfill
    \vtop{\hbox to 0in{\ \textbf{(d)}}
        \hbox{\includegraphics[width=0.30\textwidth]{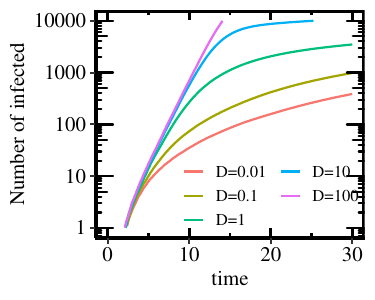}}}
    \vtop{\hbox to 0in{\ \textbf{(e)}}
        \hbox{\includegraphics[width=0.30\textwidth]{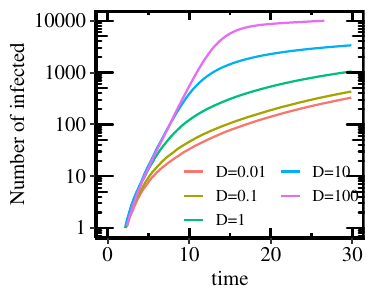}}}%
    }
    \caption{{\bf Epidemic on a population undergoing Brownian motion}. Network parameters are  $N=10^3$ nodes, $K=8$ average neighbors. The epidemic model is SI with Gamma-distributed transmission times, mean $\mu=4$, variance $\sigma^2 = 3$ and probability of infection $p_\psi = 0.9$. Panel (a): Snapshot of an epidemic with infected nodes in red and susceptible nodes in yellow. Panel (b): Epidemic growth for equal and constant diffusivity $D_0=D_1=D$ for infected and non-infected nodes. Panels (c): Epidemic growth for diffusivities $D_0=D$ for non-infected and $D_1=D/10$ for infected nodes. Panel (d): Epidemic growth for state-dependent diffusivities $D_0 = D_1 = D(1 - N_\text{inf}/N)^{100}$ where $N_\textrm{inf}$ is the number of infected nodes. Panel (e): Epidemic growth for diffusivities $D_0 = D(1 - N_\text{inf}/N)^{100}$ and $D_1 = D_0/10$.}
    \label{fig:brownian_motion}
\end{figure}

We consider a spatially-structured network in which the network evolution optionally responds to the epidemic outbreak. The temporal network is defined by the distances between $N$ diffusing Brownian particles. These particles represent individuals that move randomly and can infect each other when they are in close proximity (Fig.~\ref{fig:brownian_motion}). Specifically, particles $i=1,\ldots,N$ diffuse in two dimensions with particle-dependent diffusivity $D(i)$. Particles $i,j$ are connected by a link whenever $\| \vec{x}_i(t) - \vec{x}_j(t) \| \leq R$, where $R$ is a pre-defined contact distance and $\vec{x}_i$ is the position of particle $i$.

In the limit $D\rightarrow 0$, the number of infected individuals grows as $t^2$  due to the two-dimensional  geometry  (Fig.~\ref{fig:brownian_motion}a). In the opposite limit of large $D$, the population is well-mixed and epidemics initially grow exponentially with a rate $\Lambda$ defined by the Euler-Lotka equation $1/K = \int_0^\tau e^{-\Lambda \tau} \psi(\tau) d\tau$ \cite{grassly2008mathematical}.  We ran simulations for different constant diffusivities $D(i) = D$ to numerically explore the transition between these two regimes  (Fig.~\ref{fig:brownian_motion}b and S1 Video, S2 Video, S3 Video).

In real epidemics, individual mobility  usually depends on the current state of the epidemic. First, infected individuals might have a reduced mobility. Secondly, as the number of infected individuals grows, containment measures may limit the mobility of all individuals, regardless of whether they are infected. Our algorithm allows for the evolution of networks to depend on the current epidemic state, and can therefore be used to model these effects as well. We here present three examples. In the first, the diffusivity of infected nodes is reduced 10-fold (Fig.~\ref{fig:brownian_motion}c). In the second, the diffusivity of all nodes is scaled as $(1-N_\textrm{inf}/N)^\gamma$ where $N_\textrm{inf}$ is the number of infected individuals and $N$ the total number of individuals (Fig.~\ref{fig:brownian_motion}d). In the third example, both effects take place simultaneously (Fig.~\ref{fig:brownian_motion}e and S4 Video, S5 Video, S6 Video).

\subsubsection*{Empirical networks of instantaneous contacts}

\begin{figure}
    \centering
    \includegraphics[width=\linewidth]{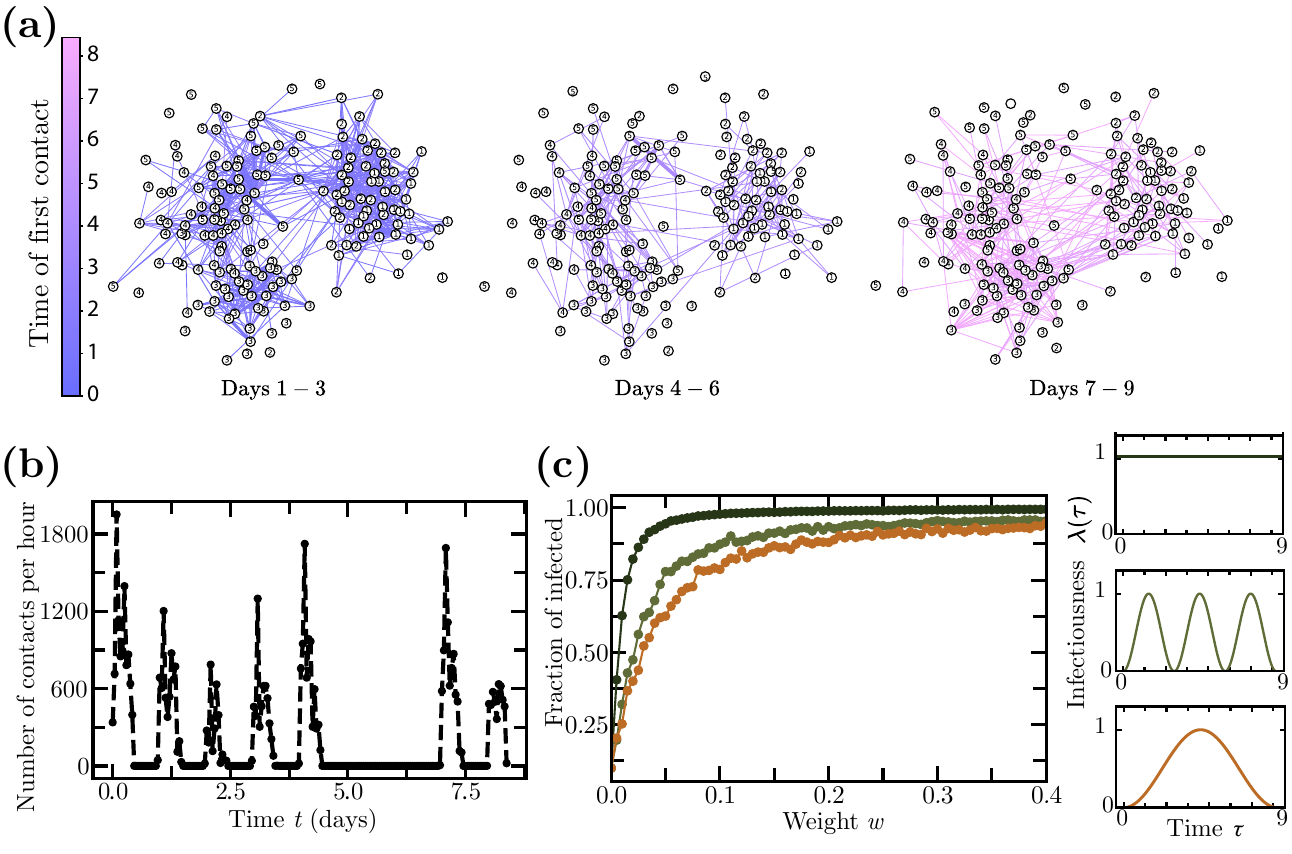}
    \caption{{\bf Epidemics on a temporal network reconstructed from face-to-face contact data in a high school in Marseille, France \cite{fournet2014contact,sociopattern}}. {\bf (a)} Illustration of the temporal network structure over nine days, comprising 180 nodes (students) and 45,047 temporal links (face-to-face contacts). {\bf (b)} Number of contacts per hour observed throughout the recorded period. {\bf (c)} Average fraction of infected individuals in simulations using different infectiousness profiles: constant infectiousness and periodic infectiousness with periods of 3 and 9 days, each peaking at maximum infectiousness $w$.  A fraction of $10\%$ of the nodes are initially infected at time $t=0$ and the epidemics are averaged over 100 simulations.}
    \label{fig:highschool_network}
\end{figure}

As an example of an empirically observed temporal network, we simulated an epidemic on a temporal network reconstructed from face-to-face contact data collected in a high school in Marseille, France \cite{fournet2014contact}. The data, obtained from Ref.~\cite{sociopattern}, include interactions among students from five classes over a span of 9 days. The resulting network comprises 180 nodes and 45047 temporal links, see Fig.~\ref{fig:highschool_network}a and Fig.~\ref{fig:highschool_network}b.

We simulate epidemics using three different choices of $\lambda(\tau)$: a constant infectiousness $\lambda(\tau)=1$, and two periodic infectiousness $\lambda(\tau)=\sin^2(\pi\tau/T)$ with periods of $T=3$ days and $T=9$ days, respectively. We run simulations for different contact weights $w$ (which effectively scale $\lambda(\tau)$) and calculate the average fraction of infected individuals in each case (Fig.~\ref{fig:highschool_network}c).

Our findings confirm that the shape of infectiousness affects epidemic spreading. Specifically, a constant infectiousness (corresponding to exponentially distributed infection times on static networks) results in a larger number of infected individuals compared to cases with periodic variations in infectiousness. 

\subsubsection*{Finite-duration vs. instantaneous contacts}

To compare the behavior and performance of instantaneous and finite-duration contacts, we tested both models on a range of different empirical contact networks \cite{panzarasa2009patterns, kumar2016edge, paranjape2017motifs, kumar2018community}.

\begin{figure}[htbp]
    \centering
    \includegraphics[width=\textwidth]{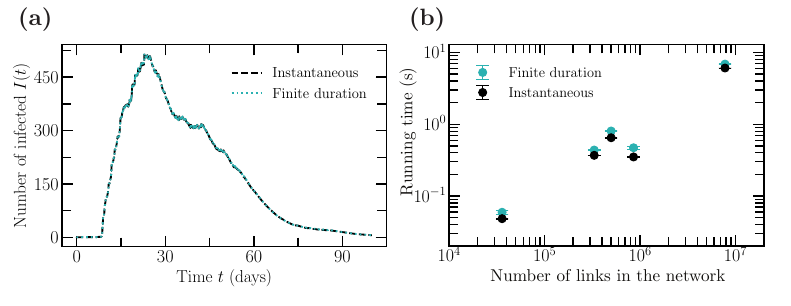}
    \caption{{\bf Empirical temporal networks with finite vs. instantaneous contacts}. {\bf (a)}:  Average trajectory over 1000 simulations of a SIR epidemic spreading along the College Messaging temporal network \cite{panzarasa2009patterns}. The dataset spans 193 days with 20296 messages, during which 1899 users have either received or sent at least one message. At the initial time, the first user is infected. For the finite-duration simulations, links with weight $w_{ij}=3$ exists for $\Delta t=10^{-5} $ days during each contact. When simulating with instantaneous contacts, contacts had weight $w_{ij}=1$. In both cases, the infectiousness $\lambda(\tau)=1$ was constant, and nodes recover after a lognormally distributed time with mean $14$ and standard deviation $10$ days. {\bf (b)}: Runtimes of an SI epidemic on different empirical temporal networks. Temporal networks were selected from an online database \cite{snapnets} and are in order of increasing size: Bitcoin web of trust network \cite{kumar2016edge}, an emails network \cite{paranjape2017motifs}, Mathoverflow \cite{paranjape2017motifs}, Hyperlinks between subreddits on Reddit \cite{kumar2018community}, User edits network on Wikipedia \cite{paranjape2017motifs}. Plot shows runtime averages over 1000 simulations.}
    \label{fig:example_empirical_temporal_network}
\end{figure}

We first compare the epidemic trajectories produced by instantaneous vs. finite-duration contacts for a SIR model of the spread of a computer virus on a network created by messages exchanged on a social networking platform at the University of California, Irvine \cite{panzarasa2009patterns} (Fig.~\ref{fig:example_empirical_temporal_network}a). Finite-duration contacts had weight $w_{ij}=1/\Delta t$ and lasted for a finite time interval $\Delta t$ centered at the reported times of contact between two nodes. Instantaneous contacts had weight $w_{ij} = 1$; this ensures that in the limit $\Delta t \to 0$, both models transmit the disease with probability $1 - \exp\big(-\lambda(\tau)\big)$ during each contact between an infected and a susceptible node. For finite $\Delta t$, we observe the resulting trajectories to be qualitatively similar, but to show some minor differences (Fig.~\ref{fig:example_empirical_temporal_network}a).

We next compared the performance for finite-duration contacts compared to instantaneous contacts for 5 other empirical networks with sizes ranging from about $3\cdot 10^4$ to about $7\cdot 10^6$ contacts (Fig.~\ref{fig:example_empirical_temporal_network}b). We find that while simulating instantaneous contacts is more efficient as we would expect, the difference in run times is only about $1.2$-fold in practice.

\section*{Discussion}

In this paper, we have presented NEXT-Net, an efficient and flexible implementation of stochastic methods to simulate epidemics on networks at the individual level. NEXT-Net includes two main algorithms: one for static networks based on the next-reaction scheme, and a newly conceived algorithm for temporal networks. Both algorithms are highly versatile, fast, and exact. In particular, the distributions of transmission and recovery times can be freely chosen, and simulations can be carried out on arbitrary weighted, unweighted, and temporal networks.

Our systematic comparisons show that NEXT-Net, besides being exact, vastly outperforms alternative methods for static networks in terms of performance. The performance gap with respect to other methods increases with increasing network size. 

For temporal networks, we are not aware of other algorithms with a similarly wide scope. The NEXT-Net algorithm can deal with a large variety of temporal network models, including the networks that react to epidemic states and which include instantaneous contacts. Despite being versatile, the algorithm still performs very well. In most practical cases, the majority of the computational time is devoted to evolving the network, rather than to the epidemic process itself. In our implementation of NEXT-NET for temporal networks, the algorithm builds on the next reaction scheme. However, the algorithm is not restricted to that, and could for example also be combined with the Gillespie or Laplace-Gillespie algorithm when the infectiousness function $\lambda(\tau)$ permits these choices of algorithms.

The algorithms in NEXT-Net are designed to be easily extensible. New transmission/recovery time distributions and static and temporal network models can be easily added without having to modify the existing algorithms. At the moment, NEXT-Net provides various types of synthetic static networks such as \ER, \BA, \WS, as well as non-clustered and clustered versions of the configuration model \cite{bender1978asymptotic,molloy1995critical,angeles2005tuning}, and allows arbitrary static networks to be defined through adjacency lists. For temporal networks, NEXT-Net comes with implementations of temporal \ER networks, activity-driven networks \cite{cai2024epidemic}, a network SIRX model \cite{maier2020a} and Brownian proximity networks, and users can add arbitrary custom models by implementing a custom time evolution procedure. In the future, we hope to further extend the range of possibilities by implementing additional models of temporal networks as they are proposed in the literature. 


NEXT-Net is available at \url{https://github.com/oist/NEXTNet} under an open-source license. To make the features of NEXT-Net easily accessible, we provide wrapper libraries for R and Python, and offer a range of empirical networks from the SNAP \cite{snapnets}, ICON \cite{clauset2016colorado} and KONECT \cite{konect} databases in a format compatible with NEXT-Net (see the NEXT-Net website). 

\section*{Acknowledgments}
We are grateful for the help and support provided by the Scientific Computing and Data Analysis section of Core Facilities at OIST.


\bibliography{bibliography}

\section*{Supporting Information}

\paragraph*{S1 Algorithms.}
{\bf Detailed description of the algorithms.}

\paragraph*{S1 Table.}
{\bf Performance data on empirical networks.} The table lists the empirical networks we analyzed, alongside their parameters and mean computing time for SIR and SIS models using different algorithms. We selected networks of size $N>1000$, only possessing undirected links, not temporal, and not bipartite, from the online databases: SNAP \cite{snapnets}, ICON \cite{clauset2016colorado} and KONECT \cite{konect}.

\paragraph*{S1 Video.}
{\bf Epidemic on a Brownian proximity network with constant low diffusivity.} Network parameters are $N=1000$ nodes, $K=8$ neighbors on average, constant diffusivities $D_0=D_1=0.01$. The epidemic model is SI with Gamma-distributed transmission times, mean $\mu=4$, variance $\sigma^2 = 3$, and probability of infection $p_\psi = 0.9$.

\paragraph*{S2 Video.}
{\bf Epidemic on a Brownian proximity network with constant medium diffusivity.} Network parameters are $N=1000$ nodes, $K=8$ neighbors on average, constant diffusivities $D_0=D_1=0.1$. The epidemic model is SI with Gamma-distributed transmission times, mean $\mu=4$, variance $\sigma^2 = 3$ and probability of infection $p_\psi = 0.9$.

\paragraph*{S3 Video.}
{\bf Epidemic on a Brownian proximity network with constant large diffusivity.} Network parameters are $N=1000$ nodes, $K=8$ neighbors on average, constant diffusivities $D_0=D_1=1$. The epidemic model is SI with Gamma-distributed transmission times, mean $\mu=4$, variance $\sigma^2 = 3$ and probability of infection $p_\psi = 0.9$.

\paragraph*{S4 Video.}
{\bf Epidemic on a Brownian proximity network with low state-dependent diffusivity.} Network parameters are $N=1000$ nodes, $K=8$ neighbors on average, diffusivities $D_0=0.01*(1 - N_\textrm{inf}/N)^{3}$ and $D_1=D_0/10$. The epidemic model is SI with Gamma-distributed transmission times, mean $\mu=4$, variance $\sigma^2 = 3$ and probability of infection $p_\psi = 0.9$.

\paragraph*{S5 Video.}
{\bf Epidemic on a Brownian proximity network with medium state-dependent diffusivity.} Network parameters are $N=1000$ nodes, $K=8$ neighbors on average, diffusivities $D_0=0.1*(1 - N_\textrm{inf}/N)^{3}$ and $D_1=D_0/10$. The epidemic model is SI with Gamma-distributed transmission times, mean $\mu=4$, variance $\sigma^2 = 3$ and probability of infection $p_\psi = 0.9$.

\paragraph*{S6 Video.}
{\bf Epidemic on a Brownian proximity network with large state-dependent diffusivity.} Network parameters are $N=1000$ nodes, $K=8$ neighbors on average, diffusivities $D_0=(1 - N_\textrm{inf}/N)^{3}$ and $D_1=D_0/10$. The epidemic model is SI with Gamma-distributed transmission times, mean $\mu=4$, variance $\sigma^2 = 3$ and probability of infection $p_\psi = 0.9$.

\end{document}


\vspace*{0.2in}

\begin{flushleft}
{\Large
\textbf\newline{Fast and exact stochastic simulations of epidemics on static and temporal networks\newline
Supplemental Information I: Algorithms} 
}
\newline
\\
Samuel Cure\textsuperscript{1},
Florian G. Pflug\textsuperscript{1},
Simone Pigolotti\textsuperscript{1*}
\\
\bigskip
\textbf{1} Biological Complexity Unit, Okinawa Institute of Science and Technology, Onna, Okinawa 904-0495, Japan.
\\
\bigskip

%
%





* simone.pigolotti@oist.jp

\end{flushleft}

\linenumbers



We here describe the NEXT-Net algorithm for static and temporal networks in greater detail and provide pseudo-code. This document is organized as follows. In Section~\ref{sec:networks}, we briefly introduce the representation of networks for the purpose of our algorithms. In Section~\ref{sec:prob}, we explain the representation of transmission and recovery time distributions. We then present the next-reaction based NEXT-Net algorithm for static networks in Section~\ref{sec:next_reaction} and discuss its computational complexity. Finally, we present the temporal NEXT-Net algorithm in Section~\ref{sec:temporal}.

\section{Networks}\label{sec:networks}

In the simulation algorithms discussed below, networks are accessed through an abstract interface which offers the following procedures: $\textsc{NetworkSize}$ returns the number of nodes in the network, $\textsc{NodeDegree}(n)$ returns the out-degree of node $n$, and $\textsc{Neighbor}(n,l)$ returns a tuple $(m,w)$ comprising the $l$-th neighbor $m$ of node $n$ and the weight $w$ of link $(n,m)$. This interface treats networks as directed graphs, meaning that it distinguishes the link $(i,j)$ connecting source $i$ to target $j$ from the link $(j,i)$ connecting source $j$ to target $i$. In our C++ implementation, this abstract interface is realized as an abstract base class. Specific types of networks such as Erdős–Rényi, Barabási–Albert, lattices and empirical networks defined by an adjacency list are implemented as separate classes, and thanks to this abstract interface can be used with all of the algorithms presented hereafter.

\section{Probability distributions}\label{sec:prob}

The NEXT-Net algorithm allows arbitrary probability distributions of a variable $\tau\in[0,\infty]$ to be used as transmission time distribution $\psi(\tau)$ and recovery time distribution $\rho(\tau)$. Infinity is explicitly included in the domains of these distributions, and represents the case of no transmission along a particular link and no recovery of a particular node, respectively. We write $\Psi(\tau)$ for the survival function associated with the density $\psi(\tau)$ assuming $w_{ij}=1$. Equation~\eqref{main-eq:psi_static} implies that
\begin{equation}\label{eq:Psi}
    \Psi(\tau) = \exp\left(-\int_0^{\tau} \lambda(\tau') d\tau'\right),
    \qquad
    \psi(\tau) = -\Psi'(\tau) = \lambda(\tau) \Psi(\tau)
\end{equation}
where $\lambda(\tau)$ is the hazard rate function which describes the instantaneous rate of transmission at time $\tau$ after infection. Any distribution can be expressed in the form of Eq.~\eqref{eq:Psi} by setting $\lambda = \psi/ \Psi$. The probability of eventual transmission along a specific link is $p_\psi = \int_0^\infty \psi(\tau) = 1 - \Psi(\infty)$.  This probability is less than one (and $\psi(\tau)$ therefore not normalized) if $\int_0^\infty d\tau\lambda(\tau) < \infty$.

To simulate epidemics on weighted and temporal networks, our algorithms requires samples from modified transmission time distributions with a shifted and scaled hazard rate function. We have introduced in the Main Text the density $\psi(\tau\,|\,w)$ for a link such that $w_{ij}=w$. Here we extend this two a two-parameter distribution $\psi(\,\cdot\,\,|\,w,t)$ defined by shifting $\lambda(\tau)$ by $t$ and scaling it by $w$. The distribution has survival function and density

\begin{align}
    \Psi(\tau\,|\,w,t) &= \exp\left(-m\int_0^\tau \lambda(t+\tau') d\tau'\right) =
    \left(\frac{\Psi(t + \tau)}{\Psi(t)}\right)^m, \\
    \qquad
    \psi(\tau\,|\,w,t) &= m \lambda(t+\tau) \Psi(\tau\,|\,w,t).
    \nonumber
\end{align}

Transmission and recovery time distributions are typically continuous, i.e. possess a continuous density. NEXT-Net also supports discrete distributions; in that case we define $\Psi(\tau)$ to be the probability that the transmission is greater or equal than $\tau$.

NEXT-Net accesses this two-parameter family of distributions over some base distribution $\psi$ through an abstract interface that  offers procedures $\textsc{DrawTime}_\psi(w,t)$ to sample from $\psi(\,\cdot\,\,|\,w,t)$ and $\textsc{HazardRate}_\psi(\tau)$ to evaluate $\lambda(\tau) = \psi(\tau) / \Psi(\tau)$. Some of the other algorithms implemented in the NEXT-Net C++ Library require some additional procedures like $\textsc{Density}_\psi(\tau,w,t)$ to evaluate $\psi(\tau\,\,|\,w,t)$, $\textsc{Survival}_\psi(\tau,w,t)$ to evaluate $\Psi(\tau\,|\,w,t)$ and $\textsc{Quantile}_\psi(p,w,t)$ to evaluate the inverse $\Psi^{-1}(\tau\,|\,w,t)$. The REGIR algorithm also requires $\textsc{HazardBound}(\tau)$ to evaluate $\max_{\tau' \in [0,\tau]} \lambda(\tau')$. Different base distributions are implemented as separate classes which implement these functions. Currently NEXT-Net provides the following distributions

\begin{description}
    \item[Exponential] Exponentially-distributed transmission time with non-normalized density $\psi(\tau) = p_\psi \lambda e^{-\lambda \tau}$, parametrized by rate $\lambda$ and probability $p_\infty = 1 - p_\psi$ of no infection. Corresponds to constant infectiousness (hazard rate) $\lambda(\tau)=\lambda$ in the case of $p_\infty = 0$.
    \item[Weibull] Weibull-distributed transmission time with non-normalized density $\psi(\tau) = p_\psi \alpha \theta^{-\alpha} \tau^{\alpha-1} e^{-(\tau/\theta)^\alpha}$, parametrized by shape $\alpha$, scale $\theta$, and probability $p_\infty = 1 - p_\psi$ of no infection. Corresponds to $\lambda(\tau) = p_\psi \alpha \theta^{-\alpha} \tau^{\alpha-1}$, i.e. an infectiousness (hazard rate) which grows/declines with exponent $\alpha-1$ in the case of $p_\infty = 0$.
    \item[Gamma] Gamma-distributed transmission time with density non-normalized $\psi(\tau) = p_\psi  \theta^{-\alpha} \tau^{\alpha-1} e^{-\tau/\theta}/\Gamma(\alpha)$ where shape $\alpha=\mu^2 / \sigma^2$ and scale $\theta=\sigma^2 / \mu$ so that the distribution is parametrized by its mean $\mu$, variance $\sigma^2$ and probability $p_\infty = 1 - p_\psi$ of no infection.
    \item[Lognormal] Log-normally distributed transmission time with non-normalized density $\psi(\tau) = e^{-(\ln \tau - m)^2(2s^2)} / \big(\tau \sigma \sqrt{2\pi}\big)$ where log-mean $m=2\log(\mu) - \log(\mu^2 + \sigma^2)/2$ and log-variance $s^2=\log(1 + \sigma^2 / \mu^2 )$ so that the distribution is parametrized by its mean $\mu$, variance $\sigma^2$ and probability $p_\infty = 1 - p_\psi$ of no infection.
    \item[Empirical infectiousness] 
    Transmission time with non-normalized density $\psi(\tau) = \lambda(\tau) \exp\left(-\int_0^\tau \lambda(\tau') d\tau'\right)$ where the infectiousness (hazard rate) is specified by vectors $(\tau_i)$, $(\lambda_i)$ with $\tau_1 \leq \cdots \tau_n$ and $\lambda_i = \lambda(\tau_i)$. Between the specified points, $\lambda(\tau)$ is interpolated linearly, for $\tau > \tau_n$, $\lambda(\tau) = \lambda_n$ so that $p_\psi < 1$ if $\tau_n = 0$.
    \item[Polynomial infectiousness] $\psi(\tau) = \lambda(\tau) \exp\left(-\int_0^\tau \lambda(\tau') d\tau'\right)$ where the infectiousness (hazard rate) is specified by a polynomial $\lambda(\tau) = c_0 + c_1 \tau + c_2 \tau^2 + \ldots$ with user-defined non-negative coefficients $c_i$.
    \item[Deterministic infection time] Deterministic transmission time $\psi(\tau) = \delta(\tau - \tau_0)$ for constant $\tau_0$.
\end{description}

\section{The next reaction method}\label{sec:next_reaction}

The next reaction method operates by maintaining a priority queue ($Q$) that always contains all future times at which an active link (that is, a link connected to an infected node) will attempt to transmit the disease. Each entry in the queue is represented by a tuple $(t,s,i,j,w)$ where $t$ is the time of the event, $s$ the type, (T) transmission or (R) recovery (in the case of SIR or SIS), $i$ the infecting node, $j$ the node that is being infected and $w$ the weight of link $(i,j)$. The algorithm is initialized with a list of initial infection times $t_1,\ldots,t_m$ of certain nodes $n_1,\ldots,n_m$ (procedure \textsc{EpidemicInit}). In addition to the priority queue $Q$, the algorithm tracks the times $T_i$ of the latest infection, and the times $R_i$ of the next recovery of node $i$; initially these times are set to $\bot$. Here, the symbol $\bot$ represents an undefined or uninitialized state.

\begin{algorithm}[htb]
\begin{algorithmic}
    \Procedure {EpidemicInit}{$(n_1,t_1),\ldots,(n_m,t_m)$}
        \State $Q \leftarrow \{(t_1,\tI,\bot,n_1,\bot),\ldots,(t_m,\tI,\bot,n_m,\bot)\}$
        \State $T_i, R_i \leftarrow \bot$ for all nodes $i$
    \EndProcedure
\end{algorithmic}
\label{proc:EpidemicInit}
\end{algorithm}

 At each step (procedure \textsc{EpidemicStep}), the algorithm retrieves the earliest event from the queue, processes it, and returns it. Transmission events cause the target node to become infected (procedure \textsc{InfectNode}) if transmission across the link is possible (procedure \textsc{TransmitAcrossLink}) and the target node is susceptible. Transmission is always possible for static networks: the procedure \textsc{TransmitAcrossLink} only blocks certain transmissions when simulating an epidemic on a temporal network, see Section \ref{sec:temporal}. Successful transmission generate further events in the queue, which are then processed by later calls to \textsc{EpidemicStep} in order of their occurence. The procedure \textsc{EpidemicStep} is iterated until either the queue is empty (at which point the epidemic has stopped) or until some stopping condition is met. 

\begin{algorithm}[htb]
\begin{algorithmic}
    \Procedure {EpidemicStep}{$t_\text{max}$}
    \If {$Q$ is empty or has no entry with time $t \leq t_\text{max}$}
        \State \textbf{return} $(\infty, \bot, \bot, \bot, \bot)$
    \EndIf
    \State fetch and remove event $(t, s, i, j, w)$ with minimal $t$ from $Q$
    \If {$s = \tI$ and $\textsc{TransmitAcrossLink}(i,j)$ and node $j$ is susceptible}
        \State $\textsc{InfectNode}(t, j)$
    \ElsIf {$s$ = \tR and SIS}
        \State mark node $j$ as susceptible
    \ElsIf {$s$ = \tR and SIR}
        \State mark node $j$ as recovered
    \Else
        \State start \textsc{EpidemicStep} from the top
    \EndIf
    \State \textbf{return} event $(t,s,i,j,w)$
    \EndProcedure
\end{algorithmic}
\end{algorithm}

Upon infection of a node $i$ (procedure \textsc{InfectNode}), the infection time $T_i$ is updated, a recovery time $R_i$ is generated (for SIR and SIS models) and all of the node's outgoing links $(i,j)$ are activated.

\begin{algorithm}[htb]
\begin{algorithmic}
    \Procedure {InfectNode}{$t, i$}
        \State mark node $i$ as infected
        \State $T_i \leftarrow t$
        \If {SIR or SIS}
            \State $R_i \leftarrow T_i + \textsc{DrawTime}_\rho(0,1)$
            \State add event $(R_i,\tR,i,i,\bot)$ to $Q$
        \EndIf
        \For {$l=1,\ldots,\textsc{NodeDegree}(i)$}
            \State $(j', w') \leftarrow \textsc{Neighbor}(i,l)$
            \State $\textsc{ActivateLink}(t, i, j', w')$
        \EndFor
    \EndProcedure
\end{algorithmic}
    \end{algorithm}

Upon activation of a link $(i,j)$ with weight $w$ (procedure \textsc{ActivateLink}), an tentative infection time $\tau$ for node $j$ is sampled and an infection event is added to the queue. For correctness on temporal networks, the time interval until infection is sampled from $\psi(\,\cdot\,\,|\,t-T_i,w)$ defined in section \ref{sec:prob}; this correctly handles links which are added retroactively after a node has already been infected. On static networks, $T_i = t$ and this condition is thus immaterial. If the infection time lies after node $i$'s recovery time, no event is added since recovered nodes cannot spread the infection.

\begin{algorithm}[htb]
\begin{algorithmic}
    \Procedure {ActivateLink}{$t,i,j,w$}
        \State $\tau \leftarrow \textsc{DrawTime}_\psi(t-T_i,w)$
        \If {$t + \tau < R_i$}
            \State add event $(t + \tau,\tI,i,j,w)$ to $Q$
        \EndIf
    \EndProcedure
\end{algorithmic}
\end{algorithm}

The next reaction algorithm also permits to query the time of the next event  without executing it, by inspecting the priority queue. This is not usually required for simulations on static networks, but it is crucial for simulations on temporal networks, see Section~\ref{sec:temporal}.

\begin{algorithm}[htb]
\begin{algorithmic}
    \Procedure {EpidemicNext}{$t_\textrm{max}$}
        \If {$Q$ is empty or has no entry with time $t \leq t_\text{max}$}
            \State \textbf{return} $\infty$
        \Else
            \State \textbf{return} time of earliest entry in $Q$
        \EndIf
    \EndProcedure
\end{algorithmic}
\end{algorithm}

\subsection{Computational complexity}\label{sec:static_complexity}

We consider an epidemic spreading on a network with basic reproduction number $R_0$, i.e. where an infected node on average causes $R_0$ subsequent infections. On such a network, we consider an epidemic with $I$ infected nodes and thus at most $|Q| = I R_0$ active links. Assuming an appropriate data-structure such as a heap is used to represent the priority queue $Q$, the time complexity of adding an entry and removing the earliest entry from $Q$ is $O(\log |Q|)$. Here, $|Q|$ denotes the number of queue entries, i.e. the number of active links. Under these assumptions the average time complexity of \textsc{EpidemicStep} is $\mathcal{O}(\log(IR_0))$.



\section{Temporal networks}\label{sec:temporal}

In the Main Text, we have defined temporal networks in terms of a function $\varepsilon_{ij}(t)$ which takes the value one if the network has a link from $i$ to $j$ at time $j$, and zero otherwise. Our algorithm adopts a computationally more efficient representation. Compared to static networks, we extend the abstract interface used to query temporal networks by two additional procedures:

$\textsc{NetworkStep}(t_\textrm{max})$ determines and executes the next change in network topology, i.e., moves to the next time at which one of the functions $\varepsilon_{ij}$ jumps. Possible changes in topology are addition of a link, removal of a link, or an instantaneous contact between nodes. The time and type of change is returned in the form of a tuple $(t,s,i,j,w)$ where $t$ is the time of change, $s$ the type ('${+}$' for an added link, '${-}$' for a removed link, '${\ast}$' for an instantaneous contact), $i$ and $j$ are the source and target nodes, and $w$ is the link's weight. Infinitesimal contacts correspond to $\delta$-peaks of $\varepsilon_{ij}$; the probability of transmission during such a contact is thus $1-\exp(-w\lambda(\tau))$. If the topology does not change by time $t_\textrm{max}$, the procedure returns no event, i.e. $\bot$. After the procedure concludes, the topology as reported by $\textsc{NodeDegree}$ and $\textsc{Neighbour}$ reflects the reported change.

$\textsc{NetworkNext}(t_\textrm{max})$ returns the time of the next event without executing the event. Calls to this procedure thus leave the topology as reported by $\textsc{NodeDegree}$ and $\textsc{Neighbour}$ unchanged, and subsequent calls to $\textsc{NetworkNext}$ report the same time until $\textsc{NetworkStep}$ (or $\textsc{EpidemicStep}$ if the network topology reacts to changes in epidemic state) is called. If no change in topology occurs until time $t_\textrm{max}$, the procedure returns $\infty$.

\subsection*{Simulating epidemics on temporal networks}

To simulate epidemics on temporal networks which may change in response to epidemic events, we rely on rejection sampling. Once a link has been activated, we do not reverse this activation before the link ``fires'', i.e., before its transmission time is reached. Instead, if a link adjacent to an infected node is removed, we mark the link as masked. When a link is masked, attempts at transmitting the disease through it are ignored. This avoids having to remove events other than the earliest one from the priority queue, which is an operation not typically supported by priority queues and likely costly. 

The simulation algorithm otherwise reuses the simulation algorithm for static networks from Section~\ref{sec:next_reaction} (or may indeed use any other simulation algorithm for which equivalent procedures \textsc{EpidemicNext}, \textsc{EpidemicStep}, \textsc{InfectNode} and \textsc{ActivateLink} can be provided). 

\begin{algorithm}[htb]
\begin{algorithmic}
    \Procedure {TemporalNext}{$t_\textrm{max}$}
        \State \textbf{return} $\min(\textsc{EpidemicNext}(t_\textrm{max}), \textsc{NetworkNext}(t_\textrm{max}))$
    \EndProcedure
\end{algorithmic}
\end{algorithm}

The network is evolved in lock-step with the simulation of the epidemic. At any time, the time of the next event is thus the earlier of two times, the time of the next epidemic event (i.e. infection or recovery) and the time of the next network event (i.e. topology change), see procedure \textsc{TemporalNext}. 

\begin{algorithm}[htb]
\begin{algorithmic}
    \Procedure {TemporalStep}{$t_\textrm{max}$}
        \If {$\textsc{TemporalNext}(t_\textrm{max}) = \textsc{NetworkNext}(t_\textrm{max}) \neq \infty$}
            \State $(t, s, i, j, w) \leftarrow \textsc{NetworkStep}(t_\textrm{max})$
            \If {node $i$ is \emph{infected}}
                \If {s = '${+}$' and link $(i,j)$ is \emph{admissible}}
                    \State $\textsc{ActivateLink}(t, i, j, w)$
                \ElsIf {s = '${+}$' and link $(i,j)$ is \emph{masked}}
                    \State update link $(i,j)$ to \emph{admissible}
                \ElsIf {s = '${-}$' and link $(i,j)$ is \emph{admissible}}
                    \State update link $(i,j)$ to \emph{masked}
                \ElsIf {s = '${\ast}$' and node $j$ is susceptible}
                    \State $p \leftarrow 1 - \exp\big(-w \cdot \textsc{HazardRate}_\psi(t-T_i)\big)$
                    \State $\textsc{InfectNode}(t, j)$ with probability $p$
                \EndIf
            \EndIf
        \ElsIf {$\textsc{TemporalNext}(t_\textrm{max}) = \textsc{EpidemicNext}(t_\textrm{max}) \neq \infty$}
            \State $(t, s, i, j, w) \leftarrow \textsc{EpidemicStep}
            (t_\textrm{max})$
            \If {$(t, s, i, j, w) = (\infty,\bot,\bot,\bot,\bot)$}
                \State start \textsc{TemporalStep} from the top
            \EndIf
            \State initialize outgoing links $(j,k)$ of infected/recovered node $j$ to \emph{admissible}
        \EndIf
        \State \textbf{return} event $(t,s,i,j,w)$
    \EndProcedure
\end{algorithmic}
\end{algorithm}

During each simulation step (procedure \textsc{TemporalStep}), the algorithm then performs either an epidemic step (similar to the static network case), or a network step (described above), depending in which time was earlier. The algorithm assigns a state to each link adjacent to an infected node: \emph{admissible} (active and may transmit, or inactive and may be activated), \emph{masked} (active, but transmissions are blocked), or \emph{transmitted} (has successfully transmitted the disease). The general procedure goes as follows: When a new outgoing link is added to an already infected node, the link is activated. When an active link is removed, it is \emph{masked}, which causes transmission attempts to be blocked. When a link is re-added while still \emph{masked}, it reverts from \emph{masked} to \emph{admissible}. When a \emph{masked} link attempts to transmit, it reverts back to \emph{admissible} to indicate that it is now inactive and must be re-activated upon being re-added. When an \emph{admissible} link transmits, it changes to state \emph{transmitted} (see Fig.~\ref{main-fig:temporal_networks} in the Main Text).

The algorithm also implements instantaneous contacts. When an instantaneous contact from an infected node to a susceptible node appears, the susceptible node is infected with probability $1-\exp(-w\lambda(\tau))$ where $\tau$ is the time since infection of the infecting node and $w$ the weight of the contact. Such events are allowed only between nodes not currently connected by a link. As an optimization to reduce memory usage, link states are stored such that links in state \emph{admissible} consume no memory.

The correct handling of \emph{masked} links during transmission attempts is ensured by \textsc{TransmitAcrossLink}. This procedure blocks transmissions across \emph{masked} links, and tracks whether links have already successfully transmitted the disease. 

\begin{algorithm}[htb]
\begin{algorithmic}
    \Procedure {TransmitAcrossLink}{$i,j$}
        \If {link $(i,j)$ is \emph{admissible}}
            \State update link $(i,j)$ to \emph{transmitted}
            \State \textbf{return} \emph{true}
        \ElsIf {link $(i,j)$ is \emph{masked}}
            \State update link $(i,j)$ to \emph{admissible}
            \State \textbf{return} \emph{false}
        \EndIf
    \EndProcedure
\end{algorithmic}
\end{algorithm}

The function \textsc{TransmitAcrossLink} is used by \textsc{EpidemicNext}, which was introduced in Section~\ref{sec:next_reaction}.

\subsection{Correctness}

We now show that the NEXT-Net algorithm indeed generates transmission times with distribution $\psi_{i,j}(\tau\,|\,w\,;\,T_i)=w\lambda(\tau)\varepsilon_{ij}(T_i+\tau) \exp\left(-w\int_0^\tau \lambda(\tau')\varepsilon_{ij}(T_i+\tau')d\tau'\right)$ stated in Eq.~\eqref{main-eq:psi_temp} in the Main Text for link $(i,j)$. 

We fix a link $(i,j)$ and first consider a modified version of the algorithm above, where \textsc{ActivateLink} is called whenever \textsc{TransmitAcrossLink} would update the state to \emph{admissible} after a blocked transmission attempt, instead of delaying re-activation until the link reappears. We observe that after this modification, the successive invocations of $\textsc{ActivateLink}$ incrementally generate tentative transmission times $\tau_1 \leq \tau_2 \leq \cdots$ (expressed relative to node $i$'s infection time) by generating waiting times $\Delta_k \sim \psi(\cdot\,|w,\tau_{k-1}\,)$ between these attempts (i.e. $\tau_{k} - \tau_{k-1} = \Delta_k$, $\tau_0 = 0$). By definition of $\psi(\cdot\,|w,t\,)$, these times are the jumps of an inhomogeneous Poisson process with intensity $w\lambda(\tau)$ in order of occurrence. In particular, the number of jumps $J(I) = |\{\tau_k\} \cap I\}|$ within $I$ thus follows a Poisson distribution with rate $\Lambda(I) = w\int_I \lambda(\tau) d\tau$ and $J(I_1)$, $J(I_2)$ are independent if $I_1$, $I_2$ are disjoint.

The (modified) algorithm rejects $\tau_k$ while $\varepsilon_{ij}(T_i + \tau_k) = 0$ and accepts the first $\tau_k$ where $\varepsilon_{ij}(T_i + \tau_k) = 1$. For notational convenience we introduce $T_\varepsilon = \{\tau\,|\,\varepsilon_{ij}(T_i + \tau) = 1\}$, and consider times $\{\tau_{k_m}\} = \{\tau_k\} \cap T_\varepsilon$ where $k_1, k_2, \ldots$ index the times $\tau_k$ with $\varepsilon_{ij}(T_i + \tau_k) = 1$ in order of occurrence. Let $J_\varepsilon(I) = |\{\tau_{k_m}\} \cap I\}|$ be the number of such times within $I$, then crucially  $J_\varepsilon(I) = J(I \cap T_\varepsilon)$. It follows immediately that (i) if $I_1$, $I_2$  are disjoint so are $I_1 \cap T_\varepsilon$, $I_2 \cap T_\varepsilon$ and hence $J_\varepsilon(I_1)$, $J_\varepsilon(I_2)$ are independent, and (ii) $J_\varepsilon(I)$ follows a Poisson distribution whose rate $\Lambda(I \cap T_\varepsilon)$ by definition of $\Lambda$ is $w\int_I \lambda(\tau) \varepsilon_{ij}(T_i+\tau) d\tau$. Therefore, $\tau_{k_1}, \tau_{k_2}, \ldots $ are the jump times of a inhomogeneous Poisson process with intensity $w\lambda(\tau) \varepsilon_{ij}(\tau)$, and $\tau_{k_1}$ is the first firing time of such a process. Consequently, $\tau_{k_1}$ is distributed according to $\psi_{i,j}(\tau\,|\,w\,;\,T_i)$. This proves the correctness of the modified algorithm.

We consider now the original algorithm where re-activation of link after a blocked transmission attempt is deferred until the link re-appears. The generated times $\tau_1 \leq \tau_2 \leq \cdots $ are not jump times of an inhomogeneous Poisson process with intensity $w\lambda(\tau)$ in this case. However, only jumps which would later be rejected because $\varepsilon_{ij}(\tau)=0$ are omitted. Therefore, the properties of $J_\varepsilon(I)$ are the same for the original and the modified algorithm, and consequently the original algorithm generates $\tau_{k_1}$ with distribution $\psi_{i,j}(\tau\,|\,w\,;\,T_i)$ as claimed.

\subsection{Computational complexity}

On temporal networks, three terms contribute to the time complexity of \textsc{TemporalStep}. The first is the time required to maintain the link states; this is $\mathcal{O}(1)$ and typically negligible. The second is the time required by \textsc{NetworkNext} to determine when the next change in network topology occurs. The complexity of \textsc{NetworkNext} depends on the specific model of temporal network. For networks such as activity-driven networks or temporal Erdös-Renyi networks, either the Gillespie algorithm or a version of the next-reaction scheme can be used to simulate the network evolution; in this case the time complexity of \textsc{NetworkNext} is either constant (Gillespie) or logarithmic in the number of active elements (next reaction). For other types of networks such as Brownian proximity networks, however, the time can scale linearly with the number of nodes.

The third contribute to the complexity of \textsc{TemporalStep} are calls to \textsc{EpidemicStep}. A single call has time complexity $\mathcal{O}\big(\log(I R_0)\big)$ as discussed in Section \ref{sec:static_complexity}. Each time \textsc{TransmitAcrossLink} encounters a \emph{masked} link and blocks transmission, an additional epidemic step has to be performed. The total contribution of  \textsc{EpidemicStep} to \textsc{TemporalStep} is thus $\mathcal{O}\big(K \log(I R_0)\big)$ where $K$ is the average number of transmission attempts required before encountering the link in state \emph{admissible}.

\nolinenumbers
